# Functional Magnetic Resonance Imaging and the Challenge of Balancing Human Security with State Security

Farhan Sahito
Institute for Software Technology
Graz University of Technology
Graz, Austria
fsahito@ist.tugraz.at

Wolfgang Slany
Institute for Software Technology
Graz University of Technology
Graz, Austria
wolfgang.slany@tugraz.at

## A. Abstract

*Recent reports reveal that violent extremists are trying to obtain insider positions that may increase the impact of any attack on critical infrastructure and could potentially endanger state services, people's lives and even democracy. It is of utmost importance to be able to adopt extreme security measures in certain high-risk situations in order to secure critical infrastructure and thus lower the level of terrorist threats while preserving the rights of citizens. To counter these threats, our research is aiming for extreme measures to analyse and evaluate human threats related assessment methods for employee screening and evaluations using cognitive analysis technology, in particular functional Magnetic Resonance Imaging (fMRI). The development of fMRI has led some researchers to conclude that this technology has forensic potential and may be useful in investing personality traits, mental illness, psychopathology, racial prejudice and religious extremism. However, critics claim that this technology may present many new human rights and ethical dilemmas and could result in potentially disastrous outcomes. The main thrust of the research is to counter above concerns and harmful consequences by presenting a set of ethical and professional guidelines that will substantially reduce the risk of unethical use of this technology. The significance of this research is to ensure the limits of the state/organisation's right to peer into an individual's thought process with and without consent, to define the parameters of a person's right to ensure that fMRI scans do not pose more than an appropriate threat to cognitive liberty, and the proper use of such information in civil, forensic and security settings.*

## B. Introduction

September 11th has marked an important turning point that exposed new types of security threats and disclosed how terrorists' pursuit of their long-term strategic objectives includes attacks on innocent civilians and critical infrastructures that could result in not only large-scale human casualties but also profound damage to national power and prestige (Birkett et al., 2011 & Sue, 2006). Recent reports[1] also reveal that violent extremists are trying to obtain insider positions in critical infrastructure. Based on these reports, it is clear that their actions pose a significant threat. States have an extreme interest in detecting malicious insiders and may in certain cases take extreme measures to assure the protection of critical infrastructure and services within their jurisdictions while preserving the rights of citizens. Despite much investigation into the motivation and psychology of malicious insiders, the fact remains that it is extremely complicated to predict insider motivation (Brancik & Ghinita, 2011). This presents operators of critical infrastructure with a dilemma to establish an appropriate level of trust w.r.t. employees.

---
1 http://info.publicintelligence.net/DHS-InsiderThreat.pdf

The purpose of our research is to alleviate these threats by focusing on a multi-layered security strategy such as training of employees, threat management, security awareness policies and employees screening. However, the main thrust of this paper is centred on extreme measures such as employee screening on critical positions using cognitive analysis technology, in particular functional Magnetic Resonance Imaging (fMRI). Proponents of this neuro-imaging technology hailed fMRI as the next "truth meter" and conclude that because of the novelty of the physiological parameters being measured, this technology may be more accurate than other traditional methods for employee screening (e.g., polygraph, see Bruni, 2012, Faulkes, 2011, McCabe, 2011 & Spence et al., 2004). The hallmark of this study is to use fMRI technology to protect critical infrastructure, by providing an acceptable level of assurance as to the integrity of individuals who have access to sensitive information or/and who require access to key assets, individuals, protectively marked state's data and material, at risk of terrorist attacks. The aim is to establish an appropriate level of trust of employees, effective monitoring and ensuring that insiders do not pose a foreseeable risk to critical infrastructure. Eliminating or reducing the likelihood of deception could lighten the burden of suspicion and mistrust to promote state security and secure human lives. Clearly this is an area of great sensitivity so we need to understand that threats to critical infrastructure are becoming increasingly frequent.

To sum up, this research examines the use of fMRI technology in critical infrastructure security. The aim of our research is to show that neuro-imaging can be an important, helpful, and successful tool for state security from an employee screening perspective. However, despite the intriguing results of many studies, there are several concerns which must be addressed prior to moving this technology to real-world application (Garnett et al., 2011). For some critics, the issues of legal, ethical and privacy violations that may clash with questions of state security and human security may raise with this technology. The significance of this research is to ensure that maintaining human security is as important as promoting state security. This paper will not discuss the methodology of implementing this technology; instead the focus will be on addressing the research challenges and related issues and to elucidate our method that includes monitoring of employee to predict or detect insider threats. We show that these methods are helpful and productive and could alleviate the burden of mistrust and increase the efficiency of threat avoidance measures. More importantly, we discuss the pros and cons of this neuro-imaging technique to ensure that both state security and human security are balanced in order to achieve the objectives of this research and that it does not lead to the conclusion that the use of this technology for employee screening is ethically dubious.

# C. Malicious Insiders in Critical Infrastructure: A Threat to State Security

Critical infrastructures are the advanced physical and cyber-based systems essential to the state's security, economic prosperity and social well-being of the nation, such as law enforcement services, power plants and information and communication services etc (Moteff, 2011). As a result of advances in technology, these critical infrastructures have become increasingly automated and interlinked. On the other side, these advances have created new vulnerabilities to physical and cyber attacks by insiders[2]. The study "Cost of Data Breach Study: United States" from 2011 reveals that insiders are the top cause of data breaches and 25 percent more costly than other types[3]. Moving data and application in IT brings with it an inherent level of risk that allows insiders to steal confidential data such as passwords and cryptographic keys, sabotage information resources as well as various types of frauds (Mathew, 2012 & Rocha et al., 2011). These threats carry debilitating impact on state's defence and economic security, including a loss of public confidence in state's services that would seriously undermine public safety and the fulfilment of key government responsibilities (Jackson, 2011).

A new report[4] issued by the US Department of Homeland Security uncovers that malicious insiders and their actions pose a significant threat to critical infrastructure in the USA and the world, and may increase the impact of any attack on critical infrastructure. According to this report, the fall edition of AQAP (a magazine published by al-Qaeda) encourages followers to use "specialized expertise and those who work in sensitive locations that would offer them unique opportunities" to conduct terrorist attacks in the world. The US authorities were stunned in 2009 in Yemen with the arrest of an alleged American recruit to al-Qaeda, Sharif Mobley, who had been employed at five different US nuclear power plants in and around Pennsylvania after successfully passing federal background checks (Sharp, 2010). The sequence of scandals induced by the 2010 as publication of classified government documents to the Wiki-Leaks website (Fenster, 2012) – in which volumes of sensitive documents were leaked by a trusted insider and ultimately published on an open website – has caused much embarrassment to the United States and other nations and represents the ultimate nightmare scenario for governments when considering the human aspect in critical infrastructure. It is indeed sobering to imagine that any organisation could fall victim to such events and the damage malicious insider can do. The US president issued an executive order in October 2011[5] to create an Insider Threat Task Force to prevent potentially damaging and embarrassing exposure of important secrets. Eugene Spafford, executive director of Purdue University's Centre for Education and Research in Information Assurance and Security, said the president's action was long overdue: "Why haven't they been doing this already? This is at least 10 years too late, if not 20[6]. " It is indeed sobering to imagine that any critical infrastructure could fall victim to such events and the damage malicious insider can do.

---

2 http://www.dhs.gov/xlibrary/assets/niac/niac_insider_threat_to_critical_infrastructures_study.pdf

3 http://www.symantec.com/about/news/resources/press_kits/detail.jsp?pkid=ponemon&om_ext_cid=biz_socmed_twitter
   _facebook_marketwire_linkedin_2011Mar_worldwide_costofdatabreach

4 http://info.publicintelligence.net/DHS-InsiderThreat.pdf

5 http://docs.govinfosecurity.com/files/external/2011wiki_eo_rel.pdf

## D. Malicious Insiders in Critical Infrastructure: Why We Cannot Stop Them?

The "insider" is an individual authorized to access an organisation's information system, network or data — based on trust (Greitzer et al., 2008). The insider threat refers to harmful acts and malicious activities that trusted insiders might carry out such as negligent use of classified data, unauthorized access to sensitive information, fraud, illicit communications with unauthorized recipients and something that causes harm to the organisation (Sun, 2011). Insiders can be system administrator, contractors, former employees, suppliers, security guards and partner employees etc. According to Noonan & Archuleta (2008) malicious insiders can be labelled as three different types of actors: 1) criminals 2) ideological or religious radicals; and 3) psychologically-impaired disgruntled or alienated employees. The motivation of malicious insider can be summarized as simple illicit financial gain; revenge for a perceived wrong; or radicalization for advancement of religious or ideological objectives (Noonan & Archuleta, 2008). Insider threats are often cited as the most serious security problem difficult to deal with as he/she has capabilities and information not known to external attackers. Governments are taking all necessary measures to swiftly eliminate any significant threat from internal vulnerabilities on their critical infrastructures as such damage would generally be catastrophic and far-reaching – such as terrorist attacks (Chiaradonna et al., 2008), but the extent to which this can be done at all is far from sufficient.

To counter human threats, agencies have invested billions of Euros in different technical measures for years now (Sarka, 2010). The current security paradigms include access control and encryption to face malicious insiders and outsiders. They are implemented through passwords, physical token authentication and biometric authentication, firewalls, encrypted data transmission, data leakage prevention, behavioural-pattern threat detection, voice stress analysis and polygraphs. Some significant techniques that are being used to mitigate particularly human factors by critical organisations include:

*1. Biometrics:* Biometrics refers to technologies that measure human body characteristics, such as voice patterns, fingerprints, DNA, retina and iris patterns, facial patterns and hand measurements, for authentication purposes. However, according to security experts this technology has not yet produced concrete results in providing scalable solutions in detecting insider and outsider threats to critical organisations and comes with an associated error probability (Parvinder, 2009). According to Jim Wayman, former director, US Government Biometrics Center, "it really isn't for security — it's for convenience"[7]. These technologies increase risks to personal privacy and security of employees with no commensurate benefit for performance. Computers are fast at computation but not very good at judgment and expert social engineers can easily fool these devices (Best, 2011).

---

6 http://www.govinfosecurity.com/articles.php?art_id=4136

7 http://www.scribd.com/doc/3099277/Why-Biometrics-is-not-a-Panacea

*2. Proximity badges:* A badge worn by an employee that can be sensed by his or her work place. A workstation might be set to lock up if an authorised user's presence is not sensed. The issue is that not all proximity badges are in fact secure. Proximity badges are a perfect attack goal for social engineers as they provide a false sense of security while being very easy to steal and/or substitute (Anderson, 2000).

*3. Access control software:* This technique is used to implement least privilege policies for users and locks a system after an idle period, requiring a password to reinstate the display. A significant insider vulnerability is the unattended, yet logged-in system. According to several studies least privilege is often difficult or costly to achieve because it is difficult to tailor access based on various attributes or constraints (Bowen, 2009).

*4. Frequent or periodic re-authentication during a user access session:* This approach is also used by various organisations to preventing an insider from masquerading as another legitimate user in the presence of a personal "token", e.g., a smart card during a session; however, various reports of breaches in the security system acknowledged that this system is not a fool-proof mechanism (Chun et al., 2011).

*5. Voice stress analysis (VSA):* Some organisations and law enforcement agencies also use voice stress analysers to determine if, e.g., a caller or employees is lying. This technology is said to record physiological stress responses that are present in the human voice in the Detection of Deception (DOD) scenario. The monetary costs are substantial: it can cost up to €20,000 to purchase VSA technology. However, several studies conducted on the reliability of computer voice analysers to detect deception showed "little validity" in the technique (Eriksson & Francisco, 2007).

*6. Polygraph:* The Polygraph method aims at determining physiological correlates of behaviour such as a set of physiological parameters. Polygraph measures the subject's psychological response by monitoring blood pressure, pulse, chest expansion and electrical conductance of the skin that mirrors the activity of the autonomic nervous system in order to detect anxiety and deception in the subject (McCabe et al., 2011). However, there have been ongoing concerns and debate over polygraph's accuracy and reliability of measurement. This conventional device also raises ethical and legal issues and the relevance of the test to the field situations (e.g., civil and judicial settings) in which it is used (Simpson, 2008 & Stern, 2002). As a result, lives may be ruined and shattered with this technology. Maher Arar, a Canadian citizen who was born in Syria is one example of a victim of the technology (Macklin, 2008). In September 2002 when he was returning to Canada with his family from Tunisia he was detained by U.S. officials while changing planes at New York airport. After 13 days of questioning with polygraph (but no court action or formal action), he was handed over to Syrian law enforcement. After torture and one year of imprisonment he was released through Canadian intervention. The Canadian government apologized to him in 2007 (after a two year study by a prestigious commission[8]) and agreed to pay him 9 million U.S. dollars.

---

8 http://www.sirc-csars.gc.ca/pdfs/cm_arar_bgv2-eng.pdf

However, the United States government has not apologized (Macklin, 2008). On the other hand, it is very easy to cheat polygraphs, and a simple internet search of polygraph counter measures can reveal many ways how to cheat this technology. One former polygrapher also charges $59.95 for his manual plus DVD offering information on beating the polygraph (Greely & Illes, 2007).

Various studies demonstrate that above devices and security software are normally designed to defend against external threats to secure critical infrastructure and do not protect against attacks aided by internal help in organisations. An insider not only has the ability to obtain or access valuable data that resides on the internal network, but he/she can obtain this data from their workstation without causing suspicion or breaking trust. With unjustified trust, people cannot keep faith in state capability if the information is not assured and safe. Christopher Porter wrote that "Security technology is not a panacea. It's a process of which technology is only a piece of the puzzle"[9]. Any security system, no matter how well designed and implemented, will have to rely on people. Technology can be used as a control, but not in isolation as it is relatively simple for a social engineer to persuade one of the critical employees to divulge their log-in details, or for a malicious insider with legitimate access to abuse his/her position. We can implement appropriate technical solutions, but we still fail to handle the human factor. According to Kevin Mitnick, one of the most notorious social engineers, the human side of security is easily exploited and constantly overlooked. Agencies spend millions of Euros on firewalls, encryption and secure access devices, and it is often money wasted when none of these measures address the weakest link in the security chain, namely people (Mitnick & Simon, 1995).

Employees at all levels of the organisation are important to the overall protection strategy for critical infrastructure. Without all employees being part of the team, the enterprise, its assets, and its employees will be open to attack from malicious insiders (Cappelli, 2009). Critcal infrastructures would not likely fill an employee position of such gravity without conducting a background investigation and constant screening as employees at critical positions may be, in some instances, the first and only line of defence, and thus vital to national security. It very frequently would be of utmost importance to adopt extreme measures to secure critical infrastructure in order to lower the level of threats while preserving the rights of citizens. To the best of our knowledge, no single current approach solves this problem.

## E. Maintaining Security using Functional Magnetic Resonance Imaging (fMRI)

To overcome above limitations, researchers recently have attempted to use brain fingerprinting or brain scanning technologies to detect insider threats. Eck (1970) argues that "Every generation has attempted to develop objective and reproducible methods to discover the truth". Similarly, due to the inherent limitations of above technologies such as polygraph, it is not surprising that research communities and intelligence personnel have started recognizing that medical science — in particular, fMRI — may have potential applications in economics contexts, justification of cognitive enhancing drugs in educational settings, detecting deception, interrogation process and be effective in courtroom situation (Garnett et al., 2011). For instance, in September 2008, a court in India allowed to use brain scan images in a criminal case. Aditi Sharma was convicted by a court for the murder of her former fiancé, Udit Bharati. However, for the first time, a brain scan was used as evidence of a criminal defendant's guilt. This case marked the dawn of a new era for the use of brain scan technology in criminal prosecution. The court found that the brain scan proved that Aditi Sharma had experimental knowledge of having murdered Udit Bharati herself (Brown & Murphy, 2009). A variety of recent advances in neurological research and the development of this new technology claim to be a more accurately deception revealing tool for screening purposes and in counterterrorism investigations, that can be effective in distinguishing truth tellers from liars and to determine hidden conscious states of an individual, with accuracy greater than chance (Faulkes, 2011 & Marks, 2007). In this research we argue that this information can be used as a tool warranted for certain extremely critical employment situations to secure key assets — as in this era of terrorism that is creating an all-pervasive fear, fMRI can be considered as a magic bullet in the war on terror (Faulkes, 2011) fMRI is an increasingly popular neuro-imaging technique that was developed in the 1990s and has since become the preferred method for studying the functional anatomy of the human brain. This technique relies on the fact that cerebral blood flow and neuronal activation are coupled. When an area of the brain is in use, blood flow to that region also increases (Simpson, 2008). This is how the fMRI detects this physiological change due to the blood-oxygen-level-dependent, or BOLD, effect. In clinical settings, fMRI has been applied ranging from language comprehension to personality traits (happiness, sadness, fear, and anger), aesthetic judgment or political behaviour (Garnett, 2011).

---

[9] http://securityblog.verizonbusiness.com/2010/12/06/security-can-not-be-addressed-by-technology-alone/

Since an initial publication in 2001 by Spence and his colleagues (Spence et al., 2001) on fMRI deception detection, several research papers and studies on the fMRI technique have reported experiments in which subjects were asked to deceive or lie in one task and respond truly in another task (Spence et al., 2001 & Nunez, 2005). In these two studies, subjects were instructed to say yes when the truth is no and vice versa (Spence et al., 2001 & Nunez, 2005). In another study, the task paradigm included spontaneous lies (Ganis et al., 2003), for instance, the subject was instructed to say Chicago when the truthful answer is Seattle. Similarly (Lee et al., 2002 & Lee et al., 2005) studies were included feigning memory impairment tasks. In addition, lying about having a play card (Langleben et al., 2002, Langleben et al., 2005, Davatzikos et al., 2005 & Phan et al., 2005) and lying about having fired a gun (Mohamed, 2006) revealed that particular spots in the brain's prefrontal cortex become more active when a subject is suppressing the truth or lying. In some of the other experimental tasks, subjects were motivated by monetary incentives as they were told that they would double their reward money if they were able to deceive the fMRI machine, such as lying about having taken a ring or a watch (Kozel et al., 2006) and lying about the place of hidden money (Kozel et al., 2004a & 2004b).

In these experiments, this technology has been claimed to be 90% accurate by these researchers. In one study, subjects were instructed to decide either to subtract or add two numbers that had been showed to them (Haynes et al., 2007). Interestingly, on the basis of fMRI technology, experimenters were able to find (with up to 70% accuracy) whether participants would subtract the presented number from the other or whether they would sum the numbers. Apart from these laboratory experiments, Sean Spence, who has pioneered the use of this groundbreaking technology, carried out a real-life experiment in 2008. He investigated the potential innocence of a woman who had been convicted of intentional inducing illness in a child (and later was sentenced to four years in prison, see Spence et al., 2008). According to Spence, this ground breaking research proves that fMRI has the potential to reduce the number of miscarriages of justice and capacity to address the question of guilt versus innocence. According to some other researchers[10], fMRI can help you to look in a person's brain to determine if he or she has been to any specific place before, so if a person was in any terrorist training camp before, you can actually determine that. To sum up, according to this and many other studies, this technology is claimed to be useful in investing personality traits, mental illness, religious extremism, racial prejudice, lie detection and employee screening.

Not only has this neuro-imaging technology taken the attention of scientific communities but it has also attracted interest of the press and corporate world (Tim, 2011), as outside the legal system fMRI has also critical importance in the insurance industry for detection of deception (Simpson, 2008). In result, two private companies, Cephos Corp and No Lie MRI, were launched in 2006 and have begun marketing their lie detection services and offering this technology with the goal of bringing these methods to the common public in legal proceedings and security investigations (Simpson, 2008).

## F. Employee Screening using fMRI: Building Trust and Improve Security

Our research is focused on functional MRI in non clinical settings, such as employee screening in critical infrastructure. Employees at critical positions need to understand that they are very important to the state's security. In the context of critical infrastructure (i.e., normally unacceptable for non-critical situations), this research is aiming at adopting extreme measures, such as employee screening for critical personnel in order to detect malicious intent activities and susceptible behaviour and other weaknesses (drug related or domestic problems and religious extremism etc.) to uncover prior criminal records, issues with character or credit problems which can help an employer assess potential risk posed by the candidate.

Employee screening is central to such an approach. It can deal with insider threats and will help to counter the full range of threats that critical organisation may face, up to and including terrorism. The Insider Threat Study has also revealed a surprisingly high number of malicious insiders with prior criminal convictions when hired[11]. Having access to a complete employee history is an effective way of performing due diligence to protect key assets. At one hand it is beneficial for a general improvement which ultimately leads to higher productivity, better workers, increased efficiency and will provide an acceptable level of assurance for employees who have access to protectively marked critical assets and could alleviate the burden of mistrust. Furthermore, the aim of introducing this scanning technique is also deterrence from malicious activity of any kind. Indeed this approach may deter some high risk candidates with criminal/terrorist backgrounds from applying for the job, which may save money and time in the recruitment process.

---

10 http://news.bbc.co.uk/nol/shared/spl/hi/programmes/analysis/transcripts/15_03_10.txt

## G. Functional MRI: Human Security and Ethical Consequences

Although above studies reported reasonably high individual accuracy rates, there are still significant legal, ethical and human security concerns must be addressed prior to moving this technology to real-world application (White, 2010 & Garnett et al., 2011). Firstly, bioethicists have for years been debating the validating and accuracy of using fMRI technology outside of a clinical setting such as civil, forensic, and security settings. According to some critics, this approach is speculative and could raise privacy debates and it is possible that this practice might be viewed as invasive and excessive by some individuals (Bizzi et al., 2009). Other argues that this technology limits the person's right to scan employee thought processes without or with his consent. It also raises concerns about the confidentiality of this information as it may violate the right to one's internal mental privacy (Wolpe et al., 2010). This could exacerbate already precarious circumstances and may lead to more severe insider threats if any screening practice might be viewed as excessive by employees (Greitzer & Frincke, 2010).

Secondly, many of the issues are directly relevant to the fMRI experts, their expertise and public responsibility as well as the transparency of ethic issues regarding the conduct of this neuro-imaging research (Choudhury et al., 2009). Thirdly, an employee who failed an fMRI test could still assert reasonable doubt in the organisation, unlike the case with DNA identification, for instance, with which the odds of being falsely recognized are on the order of millions to one (Simpson, 2008). According to Heckman & Happel (2006) fMRI has some significant disadvantage from a human security point of view. For instance, if the subject has a metallic objects in their body and is brought into the scanning room, it could be unsafe because of the strong magnetic field inside this machine. It is also risky for people with claustrophobia and pregnant women to go through the scanning process. Fourthly, organisations have to answer some critical questions regarding under what circumstances an agency should be allowed to look for screening with this technique (Marks, 2007), and finally,

---

11 Insider Threat Study: Illicit Cyber Activity in the Banking and Finance Sector.

   http://www.cert.org/archive/pdf/bankfin040820.pdf and Insider Threat Study: Computer System Sabotage in Critical

   Infrastructure Sectors. http://www.cert.org/archive/pdf/insidercross051105.pdf.

it remains an open question how well employee screening with fMRI technology would work to ensure that human security is considered as important as state security.

## H. Functional MRI & the Dynamics between Human and State Security

According to White (2011) ethical conflicts often arise when clinical technology is used for non-clinical purposes. However, it should be clear that fMRI is not a mind reading technology (Schweitzer, 2011). According to Jones (2009) this technique does not provide any precise conclusion about a person's thought or what a person is thinking. It can only show a difference across time, across location and across tasks. This technique is very good at discovering when brain tissues are active during different cognitive tasks. This strongly suggests that fMRI does not violate the right to internal mental privacy. Secondly, White (2010) pointed out that fMRI is ethically acceptable in the market to the same extent as traditional polygraphs, and if clients are permitted to undergo a traditional polygraph examination in employee screening, the argument is equally strong concerning fMRI scans.

However, the human security issues raised by critics are complex and it is possible that this technology may be misused by some organisations. The challenge, therefore, is to forge a consensus on balancing the pursuit of human and state security to protect critical infrastructure. Consequently our aim in this research is to find a good combination of ethical guidelines that could ultimately become a general method for employee screening in critical situations and conversely decrease the extent to which it is misused or misunderstood:

1. Informed consent should be sought before fMRI scanning as the employee should be aware of the potential dangers and he/she should read, understand and sign an informed consent disclaimer to ensure that all the necessary requirements are met.
2. fMRI scan should not be allowed and should be unconstitutional unless it is done with the informed consent of the employee (Simpson, 2008). An employee who undergoes an fMRI scanning process must not be harmed by incidental findings.
3. Any pre-employment screening process must be compatible with all relevant legislations, for instance, Human Rights legislation. Question should be limited to a verification of the "real" or "personal" identity such as education, employment history, court records, credentials and other data associated with an employee.
4. In post-employment screening a policy can be introduced of only screening in case of suspicious activities. Drug testing policies[12] of the Österreichischer Gewerkschaftsbund (ÖGB) in Austria, Deutscher Gewerkschaftsbund (DGB) in Germany, and the Confédération Générale de Travail (CGT) in France are suitable case studies for this approach.
5. There must be greater regulatory controls in place to protect employees during fMRI scans including the required training of fMRI operators. Operators should be required to participate in training and receive certification and only trained experts are required to evaluate employees and conduct the scan. Secondly, an accrediting body should certify fMRI facilities. It must ensure that employees with metal plates or screws in their bones, pregnant women and claustrophobia patients should not be scanned (Rosen & Gur, 2002).
6. If an expert does not detect any abnormal behaviour, the employee is not harmed. However, if an abnormality is detected, the results of the scan should be analysed by at least another highly trained expert and possibly rectified.
7. To assure the safety of the employee, the fMRI scan process should undergo a complete governmental approval process to make reasonable assurance of employee's safety.
8. Employee's right must be protected by Article 8[13] of the European Convention on the Protection of Human Rights and Article 12[14] of the Universal Declaration of Human Rights. For instance, Article 8 guarantees the right to privacy, except "in the interest of national security, public safety or the economic well-being of the country, for the prevention of disorder and crime, for the protection of health and morals, or for the protection of the rights and freedoms of others".
9. fMRI scanning must ensure the privacy issue links to the question of data protection. The scan must implement the United Nations International Labour Organisation (ILO) code of practice on the Protection of Workers' Personal Data (1996)[15] as well as European Union Guidelines 95/46 and 97/66 on data protection.
10. A critical organisation that dismisses an employee for failing an fMRI scan test must be able to justify the action against him/her under the influence of a Human Rights Act such as the European Convention on Human Rights (ECHR)[16] or the UK Human Rights Act 1998[17].
11. Government should ban fMRI scanning in relation to non critical organisation and should also ban any non-research use of fMRI scan until it is approved by a regulatory agency.

---

12 http://www.alcoholdrugsandwork.eu/resources/ilo-ethical-issues-in-workplace-drug-testing-in- europe.pdf

# I. Conclusion

In this research, we examined the use of fMRI in critical infrastructure to detect insider threats and conclude that this brain imaging technology can be an important, helpful, and successful tool for maintaining state security, as it may provide a more reliable method of getting and evaluating information from individuals. Furthermore, the method of dealing with employees should start in the hiring process. A consistent practice of performing background checks and evaluating individuals based on the information (such as past employment, previous criminal convictions, drug related problem and verify credentials) obtained through fMRI can reduce insider threats in critical infrastructures. However, there are ethical and legal concerns that must be considered before employing this technique. These proactive measures should not be punitive in nature; rather, the employees should be educated about them with appropriate care (Cappelli et al., 2009). Our guidelines and research show that human security should be considered as equally important as the security of state. However, employees' working at critical positions must understand that under certain circumstances such as terrorism threats, the security of the state takes precedence, and respect for national sovereignty must prevail, as security of the whole nation may be threatened. However, if the fMRI scanning process is fully disclosed, explained and managed equitably, it is not as likely to be considered unfair by employees in critical positions and the mutual trust relationship required for an organisation is more likely to remain intact. The fMRI scanning process then becomes a known and understood element of the conditions of employment. Furthermore, government regulation appears to be a good way to accomplish this milestone. Our research is a first step towards maximizing the benefits of this emerging technology while minimizing the harms. It is our conclusion that the use of fMRI for employee screening can be accepted under the condition of informed consent. However, the best and first line of defence is a commitment by organisations to ensure that insofar as it is possible, its employees are satisfied, engaged and treated fairly.

---

13 http://www.echr.coe.int/NR/rdonlyres/D5CC24A7-DC13-4318-B457-5C9014916D7A/0/ENGCONV.pdf

14 http://www.un.org/en/documents/udhr/index.shtml#a12

15 http://www.ilo.org/wcmsp5/groups/public/---ed_protect/---protrav/---safework/documents/normativeinstrument/wcms_107797.pdf

16 http://www.echr.coe.int/NR/rdonlyres/D5CC24A7-DC13-4318-B457-5C9014916D7A/0/ENG_CONV.pdf

17 http://www.legislation.gov.uk/ukpga/1998/42/contents

## J. References: